\documentclass[aps,prx,superscriptaddress,nolongbibliography,twocolumn,notitlepage,nofootinbib,10pt]{revtex4-2}
\usepackage{xcolor}
\definecolor{color1}{rgb}{0,0,0.7}
\definecolor{color2}{rgb}{0.85,0,0}

\usepackage[breaklinks=true,colorlinks,citecolor=red,linkcolor=color2,urlcolor=color1]{hyperref}
\usepackage{graphicx}
\usepackage{amsmath}
\usepackage{amssymb}
\usepackage{amsfonts}
\usepackage{amsthm}
\usepackage{bbm}
\usepackage{hyperref}
\usepackage{soul}
\usepackage{textcomp}
\usepackage{bm}
\usepackage{dsfont}
\usepackage{relsize}
\usepackage{braket}
\usepackage{enumitem}
\usepackage{cleveref}
\usepackage{bigints}
\usepackage{float}
\usepackage{framed}
\usepackage[makeroom]{cancel}

\usepackage{pifont}

\usepackage{titlesec}
\titlespacing*{\section}{0pt}{10pt plus 4pt minus 5pt}{7pt plus 3pt minus 5pt}
\titlespacing*{\subsection}{0pt}{5pt plus 3pt minus 1pt}{4pt plus 2pt minus 4pt}
\titleformat{\section}{\centering\bfseries}{\thesection.}{.5em}{}

\def\Tr{{\rm Tr}}

\newcommand{\eref}[1]{\textcolor{color2}{\hyperref[#1]{eq.$\,$(\ref{#1})}}}
\newcommand{\Eref}[1]{\textcolor{color2}{\hyperref[#1]{Eq.$\,$(\ref{#1})}}}
\newcommand{\fref}[1]{\textcolor{color2}{\hyperref[#1]{Fig.$\,$\ref{#1}}}}

\newcommand{\sfref}[2]{\textcolor{color2}{\hyperref[#1]{Fig.$\,$\ref{#1}(#2)}}}
\newcommand{\tref}[1]{\textcolor{color2}{\hyperref[#1]{Table~\ref{#1}}}}

\newcommand{\aref}[1]{\textcolor{color2}{\hyperref[#1]{App.$\,$\ref{#1}}}}
\newcommand{\sref}[1]{\textcolor{color2}{\hyperref[#1]{Sec.$\,$\ref{#1}}}}

\def\eps{\varepsilon}

\DeclareMathOperator*{\argmin}{arg\,min}

\def\Tr{{\rm Tr}}

\newcommand{\nocontentsline}[3]{}

\titlespacing*{\section}{0pt}{10pt plus 4pt minus 2pt}{5pt plus 0pt minus 5pt}

\begin{document}

\title{Optimal Finite-Time Thermodynamics of Effective Two-Level Systems}

\author{Alberto Rolandi}
\email{alberto.rolandi@tuwien.ac.at}
\affiliation{Atominstitut, TU Wien, 1020 Vienna, Austria}
\affiliation{Institute for Quantum Optics and Quantum Information - IQOQI Vienna, Austrian Academy of Sciences, Boltzmanngasse 3, A-1090 Vienna, Austria}

\begin{abstract}
	The optimization of the conversion of thermal energy into work and the minimization of dissipation for nano- and mesoscopic systems is a complex challenge because of the important role fluctuations play on the dynamics of small systems. We generalize the work of Esposito et al. EPL 89, 20003 (2010) to optimize at all driving speeds the control needed to extract the maximum amount of work from any effective two-level systems. These emerge when one coarse-grains degrees of freedom, which is often unavoidable to obtain ``real-world'' two-level systems. In particular, we allow even for the system to have underlying quantum dynamics, as long as these allow for a coarse-graining that leads to a Markovian master equation. We analyze the finite-time thermodynamics of these systems and find the thermodynamically optimal protocols, which depend on the size of the coarse-graining needed to obtain a two-level system. Furthermore, we use these results to derive speed-limits for any transformation performed on an effective two-level system.
\end{abstract}

\maketitle

\section{Introduction}
The optimization of the work cost of physical tasks is a paradigmatic topic in thermodynamics, dating back to the very inception of the field~\cite{Carnot1824}. Beyond its self-evident applicative aspects ranging from macroscopic to microscopic scales~\cite{Carnot1824,Fermi1956,Myers2022,McMahon2023}, this question also probes the fundamental thermodynamic cost of irreversibility~\cite{Nernst1906,Kox2006,Benenti2017,roadmap2025}. The finite-time optimization of work has been extensively studied in systems close to equilibrium~\cite{Schmiedl2007,Salamon1983,Sivak2012,Holubec2016,Dechant2019,Deffner2020,VanVu2021,Andresen2022,Rolandi2023Collective,VanVu2023,roadmap2025,Rolandi2023Quantum}. Yet, small systems--such as biological molecules or nanoscale devices--can easily be taken far from their equilibrium state, where they show large fluctuations in their behavior. In these regimes, fluctuation theorems~\cite{Jarzynski1997,Crooks1999} provide a very powerful machinery to study systems far from equilibrium in the stochastic setting. Still, many open questions remain, particularly when quantum dynamics cannot be neglected~\cite{Schmiedl2007,Esposito2010a,Aurell2011,Benenti2017,Miller2019,Baeumer2019,Ye2022,Erdman2023a,roadmap2025,Bonanca2018}.


Two-level systems play an central role in the study of thermodynamic optimization in the mesoscopic regime as they reduce the problem to its most fundamental components~\cite{Allahverdyan2013,Abiuso2020,Denzler2021}. However, most often the optimization problem remains still too complex to solve it in full generality.
In this work, we adopt an optimal control framework to generalize the pioneering work of Ref.~\cite{Esposito2010} to \emph{effective} two level systems, where each level may be arbitrarily degenerate. This generalization allows for the implementation of this optimization procedure into a wide variety of nano- and mesoscopic two-level systems, for arbitrary driving speeds. In realistic settings, such ``two-level'' systems often arise by coarse-graining multiple degenerate states into a single effective energy level. The resulting system has thermodynamic and dynamical properties that are significantly different from those of the microscopic case where there are no degeneracies. Therefore these differences must be taken into account when studying the out-of-equilibrium dynamics of these effective systems. A simple example of such a system is a quantum dot where the two logical states are defined via electronic occupation: the empty state is non-degenerate, while the occupied state exhibits twofold degeneracy due to the spin of the electron~\cite{Aggarwal2025}. Furthermore, these different properties that effective two-level systems exhibit can also be desirable.  For instance, they emerge naturally as the energy structure that maximizes the sensitivity of thermometers~\cite{Correa2015,Mehboudi2022,Abiuso2024} and to approach reversibility in thermodynamic operations~\cite{Rolandi2023Collective,Liang2025}.

Here we consider scenarios where one can extract ($W<0$) or inject ($W>0$) some amount of work $W$ from/into the system by changing the energy gap $E$ between the two coarse-grained levels in some finite amount of time $\tau$, while it interacts with a thermal reservoir. The amount of work can be split into a reversible and irreversible contribution $W = \Delta F_{neq} + W_{diss}$. $\Delta F_{neq}$ is the non-equilibrium free energy difference, and depends exclusively on the initial and final conditions of the state and Hamiltonian. Whereas $W_{diss}$ is the dissipated energy, which depends on the way the energy gap is changed and the state trajectory throughout the process. By the second law of thermodynamics, $W_{diss}\geq 0$, with equality in the reversible limit. For fixed boundary conditions, the task of minimizing the work cost is equivalent to minimizing the dissipated energy. In general this is a hard objective because of the high degree of complexity in the dynamics of open quantum systems. 
Multiple approaches exist to obtain analytical results. For example, one can make assumptions on the driving speed of the Hamiltonian: either that the system does not have the time to evolve during the process~\cite{Erdman2019,Cavina2021,Blaber2021,Rolandi2023Fast}, or that it remains close to its steady state~\cite{Scandi2019,Scandi2020,Abiuso2020entropy,Mehboudi2022,Rolandi2023Fast,Lacerda2025,Scandi2025}. Other approaches, such as optimal transport theory~\cite{Pietzonka2018,VanVu2023,Beatty2025,Cavina2018optimal,Koyuk2018,Gu2023} and thermodynamic control~\cite{Esposito2010,Bonanca2018,Cavina2018variational,Deffner2020,Aggarwal2025} allow us to also obtain analytical results by making different sets of assumptions. Here, similarly to \cite{Esposito2010,Aggarwal2025}, we restrict the system dynamics to a family of master equations, allowing us to obtain exact results on the optimal driving of $E(t)$ without any assumption on the driving speed.
\vfill

\section{Framework}
We consider an open (finite-dimensional) quantum system in contact with a thermal bath, where its energetic structure is equivalent to that of a two-level system $\hat H(t) = E_e(t)\hat\Pi_e + E_g(t)\hat\Pi_g$, where $\hat\Pi_e$ is a rank $n$ projector and $\hat\Pi_g$ is a rank $m$ projector, with $\hat\Pi_g\hat\Pi_e = \hat\Pi_e\hat\Pi_g = 0$. These projectors define what we refer to as the \emph{excited} sector $\mathcal H_e$ and the \emph{ground} sector $\mathcal H_g$, so that we can decompose the Hilbert space of the system as a direct sum $\mathcal{H} = \mathcal H_g\oplus\mathcal H_e$. Defining the energy gap as $E(t) := E_e(t) - E_g(t)$ and denoting the state of the system at time $t$ as $\hat\rho(t)$, we can express the expected work cost of a transformation as
\begin{equation}
	W = \int_0^\tau\!dt~\dot E(t)\Tr[\hat\Pi_e\hat\rho(t)]~.
\end{equation}
We want restrict our attention to cases where each sector can be coarse-grained into a single effective state, while preserving the Markovian dynamics in the coarse-grained description. This cannot be done in general, even if we start from a fully stochastic description. As we will see below, the main assumption required for this coarse-graining is that the jump rates and operators that have support across the sectors do not distinguish the states within a sector. It is worth noting that this assumption does not impose strong restrictions on the quantum dynamics within each sector. Therefore, in the interest to keep the description as general as possible, we start from a Lindblad master equation with one bath at inverse temperature $\beta$\footnote{Here we are not including chemical potentials in the description, because we are refraining from assigning a meaning of particle number to the energy levels. However, if we were to assign $\mathcal H_e$ and $\mathcal H_g$ to different Fock space sectors, then we can introduce a chemical potential $\mu$ to the thermal bath. By construction the Hamiltonian commutes with the particle number operator, and therefore we can reabsorb $\mu$ in the energy gap $E(t)$ since there is only one bath present.}
\begin{equation}
	\frac{d}{dt}\hat\rho(t) = -i[\hat H(t),\hat\rho(t)] +\sum_{\omega\in\Omega} f(-\omega) \mathcal D_{\omega}[\hat\rho(t)]~,
\end{equation}
where $\Omega=\{-E(t),0,E(t)\}$ are the system's energy gaps, $f(\omega) = (1+e^{\beta\omega})^{-1}$ is the Fermi distribution, and the dissipator is given by
\begin{equation}
	\mathcal D_{\omega}[\hat\rho] = \sum_{i,j}\hat L_i(\omega)\hat\rho \hat L_j(\omega)^\dagger - \frac{1}{2}\{\hat L_j(\omega)^\dagger \hat L_i(\omega),\hat\rho\}.
\end{equation}
The operators $\hat L_i = \sum_{\omega\in\Omega}\hat L_i(\omega)= \sum_{\omega\in\Omega}\hat L_i(\omega)^\dagger$\footnote{$\hat L_i(E(t)) = \hat\Pi_g\hat L_i\hat\Pi_e$, $\hat L_i(-E(t)) = \hat\Pi_e\hat L_i\hat\Pi_g$,\\ $\hat L_i(0) = \hat\Pi_g\hat L_i\hat\Pi_g + \hat\Pi_e\hat L_i\hat\Pi_e$.} are the jump operators, which originate from the description of the microscopic interaction between system and bath~\cite{Esposito2006,Breuer2007}. For $\ket{1}\in\mathcal H_e$ and $\ket{0}\in\mathcal H_g$ the stochastic transition rates between the states are given by
\begin{equation}
	\Gamma_{\!1\rightarrow 0} = \gamma_{1,0}(1-f(E(t)))~,
	\quad\Gamma_{\!0\rightarrow 1}= \gamma_{0,1}f(E(t))~,
\end{equation}
for $\gamma_{1,0} = \gamma_{0,1}  = \sum_{i,j}\braket{1|\hat L_i|0}\!\braket{0|\hat L_j|1}$.

To allow for a Markovian coarse-graining, we first assume that the state contains no coherences between the ground and excited sectors $\hat\rho(t) = \hat\rho_{g}(t)\oplus\hat\rho_{e}(t)$. Second,
we assume that the jump operators induce transitions between sectors at uniform rates: $\gamma = \gamma_{1,0}$, $\Gamma_{\!e\rightarrow g} = \Gamma_{\!1\rightarrow 0}$, and $\Gamma_{\!g\rightarrow e} = \Gamma_{\!0\rightarrow 1}$ for all $\ket{1}\in\mathcal H_e$ and $\ket{0}\in\mathcal H_g$. Microscopically, this implies that the couplings to the bath that span over both energy sectors do not distinguish the internal structure of the system within an energy sector. It is worth noting that this assumption does not affect jump operator that can be written as a direct sum over the two sectors, as these do not affect $\gamma_{0}$ and $\gamma_{1}$.
While the Hamiltonian evolution is trivial since $[\hat H(t),\hat\rho(t)] = 0$, these jump operators could lead to non-trivial quantum dynamics within each sector.

This assumption is relevant for physical systems that have internal degrees of freedom which are irrelevant to the dynamics across energy levels, while still allowing for arbitrary effects within each energy sector. A natural scenario where this property can emerge is when one assigns $\mathcal H_e$ and $\mathcal H_g$ to be different Fock sectors (i.e. particle number sectors of the Hilbert space).
For example, when one encodes a two-level system in terms of photon number in a time-bin encoding, $\hat\Pi_e$ and $\hat\Pi_g$ become projectors over different Fock space sectors, which have internal degrees of freedom in the form of the photon polarization~\cite{Boaron2018,Williams2021}. Another relevant example would be when one encodes a quantum dot with the presence (or absence) of an electron. The spin degeneracy then provides an additional internal degree of freedom in the excited (or ground) sector~\cite{Aggarwal2025,Scandi2022,Barker2022}.\\

At this point we can coarse grain the sectors to a single state each by defining the probability of excitation to be $p(t) := \Tr[\hat\Pi_e \hat\rho(t)]$, so that the ground state probability is simply $1-p(t)$. The assumptions above are the minimal ones that guarantee that the coarse grained probability still follows a Pauli master equation $\dot p(t) = \Gamma_{g\rightarrow e}(1-p(t)) -\Gamma_{e\rightarrow g}p(t)$~\cite{Breuer2007}, which becomes
\begin{equation}\label{eq:deg_master_eq_0}
	\dot p(t) = \gamma \Big(n f(E(t)) - \left[m+(n-m)f(E(t))\right]p(t)\Big)~,
\end{equation}
for $n = \Tr[\hat\Pi_e]$ and $m=\Tr[\hat\Pi_g]$. The presence of the degeneracies modifies the thermal steady state probability from $f(E)$ to
\begin{equation}\label{eq:equilibrium}
	p_{eq} = \frac{1}{1+re^{\beta E}}~,
\end{equation}
where we defined the \emph{degeneracy ratio} $r=\frac{m}{n}$. It is worth noting that this immediately shows how the presence of degeneracies changes non-trivially the thermodynamical aspects of the system when compared to a true two-level system ($n=m=1$). In particular, \eref{eq:equilibrium} shows that the equilibrium properties are modified whenever $r\neq1$. Despite that, we can still emulate the equilibrium properties of a true two-level system with energy gap $\eps$ with an effective two level system by introducing an energy bias $E = \eps - k_BT\ln r$. However, while the equilibrium properties can be emulated, the dynamics are fundamentally changed. Using that $\dot p_{eq} = 0$ we can rewrite \eref{eq:deg_master_eq_0} as
\begin{equation}\label{eq:to_equilibrium}
	\dot p(t) = n\gamma\left[r+(1-r)f(E(t))\right](p_{eq}-p(t))~.
\end{equation}
In the case of a true two-level system we would have $\dot p(t) = \gamma(p_{eq}-p(t))$. When the degeneracy ratio is equal to $1$ the only difference lies in a re-scaling of the thermalization timescale $\gamma$. For $r\neq 1$, however, the result it a thermalization rate that is dependent on the energy gap. For systems where the energy gap is not driven, this is also equivalent to a simple time-scale re-scaling. However, for driven systems this re-scaling breaks down as it becomes time-dependent and it changes the dynamics in a fundamental way---in particular for optimizations and therefore the derivation of fundamental bounds. For example, in the limit of $r\rightarrow 0$ we observe a critical slowdown of thermalization for large energy gaps. 

In what follows, we will use time units such that $n\gamma=1$ to lighten the notation. Then \eref{eq:deg_master_eq_0} becomes
\begin{equation}\label{eq:master_eq}
	\dot p(t) = f(E(t)) - \left[r+(1-r)f(E(t))\right]p(t)~.
\end{equation}
The thermodynamic optimization of this type of systems has already been studied in the literature for specific degeneracy ratios. In particular, Ref. \cite{Esposito2010} covers the case $r=1$ and Ref.  \cite{Aggarwal2025} treats $r=\frac{1}{2}$.

\section{Thermodynamic Optimization}
The thermodynamic optimization of a protocol is the functional minimization of the work functional $W[p(t),E(t)] = \int_0^\tau \!dt~\dot E(t)p(t)$ for some given (arbitrary) boundary conditions on both the probability and the energy.
Via \eref{eq:master_eq} we can express the energy $E(t)$ as a function of $p(t)$ and $\dot p(t)$, therefore we can express the optimization problem as a Lagrangian extremization. Following the same steps as in \cite{Esposito2010} and \cite{Aggarwal2025}, we integrate the work by parts to avoid having to express $\dot E(t)$ in terms of $p(t)$ and $\dot p(t)$
\begin{equation}
	W = \Delta E - \int_0^\tau\!dt~ E(t)\dot p(t)~,
\end{equation}
where $\Delta E = E(\tau)p(\tau)-E(0)p(0)$ and the second term (including the minus sign) corresponds to the heat. Since $\Delta E$ is set by the boundary conditions, minimizing the work cost is equivalent to minimizing the dissipated heat. Inverting \eref{eq:master_eq} we find that we can express the energy gap as 
\begin{equation}\label{eq:energy_p_pdot}
	E(t) = k_B T\ln\!\left[\frac{1-p(t)-\dot p(t)}{\dot p(t) + rp(t)}\right]~.
\end{equation}
Therefore the Lagrangian to extremize is
\begin{equation}
	\mathcal L[p(t),\dot p(t)] = \dot p(t)\ln\!\left[\frac{1-p(t)-\dot p(t)}{\dot p(t) + rp(t)}\right]~.
\end{equation}
Since this Lagrangian is not explicitly time-dependent, by Noether's theorem, its extremizer will conserve the quantity $K = \mathcal L[p,\dot p] - \dot p\frac{\partial \mathcal L[p,\dot p]}{\partial\dot p} = \dot p^2 \frac{1- (1-r)p }{(1-p-\dot p)(\dot p + rp)}$ along the whole trajectory of $p(t)$ and $\dot p(t)$~\cite{Noether1918}. By solving for $\dot p$ we obtain
\begin{equation}\label{eq:p_dot}
	\dot p = \frac{1}{2} \frac{K\left[1-(1+r)p\right] \pm \sqrt{\Delta}}{1-(1-r)p + K}~,
\end{equation}
with $\Delta = K^2\left[1-(1-r)p\right]^2 + 4Krp(1-p)\left[1-(1-r)p\right]$. The $\pm$ sign matches the sign of $\dot p$, which is determined by the boundary conditions: if $p(\tau)-p(0)>0$ ($<0$) then we take the $+$ ($-$) sign. Interestingly, in \eref{eq:p_dot} the time and probability are separable, therefore we can integrate to obtain
\begin{equation}\label{eq:t_of_p}
	t = 2\int_{p(0)}^{p(t)}\!dp\frac{1-(1-r)p + K}{K\left[1-(1+r)p\right] \pm \sqrt{\Delta}}~.
\end{equation}
By defining $F_K(p)$ as the integrand\footnote{Since it is defined up to a constant we set $F_K(p(0)) = 0$.} of \eref{eq:t_of_p} we obtain a solution for the optimal trajectory in terms of an inverse function $p(t) = F_K^{-1}(t)$. With the boundary conditions, \eref{eq:t_of_p} sets the integration constant to $K=\kappa_\tau$, which is defined by the following equation
\begin{equation}\label{eq:K_opt}
F_{\kappa_\tau}(p(\tau)) = \tau~.
\end{equation}
It is therefore a function of $p(0)$, $p(\tau)$, and $\tau$ only.

At this point, with the integration constant set by \eref{eq:K_opt}, the optimization problem is solved. It is worth noting that applications can be done efficiently as it requires solving three scalar equations numerically\footnote{Except for a few special values of $r$ \eref{eq:t_of_p} cannot be integrated analytically. Furthermore, even when it is, \eref{eq:K_opt} cannot be solved analytically.} instead of a boundary value problem. By combining \eref{eq:energy_p_pdot} and \eref{eq:p_dot} we can obtain the optimal control of the energy gap as a function of the probability.
\begin{equation}\label{eq:energy_opt}
	E_{\kappa_\tau}(p) = k_BT\ln\!\left[\frac{\left[1-(1-r)p\right] \left(\kappa_\tau+2(1-p)\right) \mp \sqrt{\Delta}}{\left[1-(1-r)p\right] \left(\kappa_\tau+2rp\right)\pm \sqrt{\Delta}}\right]~.
\end{equation}
Therefore the optimal driving as a function of time is given by $E_{opt}(F_{\kappa_\tau}^{-1}(t))$ and the minimal work cost can be expressed as function of the boundary conditions
\begin{equation}\label{eq:work_opt}
	W_{min} = \Delta E - \int_{p(0)}^{p(\tau)}\!dp~E_{\kappa_\tau}(p)~.
\end{equation}
It is interesting to note that in the above analysis the only place where the boundary condition on the energies enters is in the work cost, while the bulk of the optimal protocol and heat exchanged are completely independent of it. Therefore the optimal protocol will require to jump to $E_{\kappa_\tau}(p(0))$ (from $E_{\kappa_\tau}(p(\tau))$) at $0^+$ ($\tau^-$). This remains true even when we assume the system to start and end at equilibrium: from \eref{eq:equilibrium} we have that $E = k_BT\ln\!\frac{1-p}{rp}$ at equilibrium, which does not match \eref{eq:energy_opt}. While the system is being driven it is ``lagging behind'' its instantaneous equilibrium state, as it can be seen from its trajectory, which we obtain by inserting \eref{eq:master_eq} into the definition of $K$
\begin{equation}\label{eq:optimal_prob}
		p(t) = p_{eq}(t)\!\left[1-A(t)\mp\sqrt{A(t)^2+r\kappa_\tau\frac{1+e^{\beta E(t)}}{r+e^{-\beta E(t)}}}\right],
\end{equation}
where we made the time-dependence of the driven equilibrium explicit and defined  $A(t)=\frac{\kappa_\tau}{2} \frac{1-r}{r+e^{-\beta E(t)}}$\footnote{In the limit of $r\rightarrow 1$ we recover the result of \cite{Esposito2010}.}. Therefore, when making an equilibrium-to-equilibrium transformation the optimal protocol will overshoot the final value of the energy to then quench back to it.\\

Before moving on to the study of of limiting cases and a few numerical examples, let us note that not all scenarios impose boundary conditions on the probabilities. It is often the case that one wishes to drive only the system's energy to some specific value $E(\tau)$ and let it relax to equilibrium after the driving. This allows for one more layer of optimization and removes the constraint imposed by \eref{eq:K_opt}. Here the constant $\kappa_\tau$ is determined by minimizing directly \eref{eq:work_opt}
\begin{equation}\label{eq:no_bound_min}
	\kappa_\tau = \argmin_{K} F_K^{-1}(\tau)E(\tau) - \int_{p(0)}^{F_K^{-1}(\tau)}\!dp~ E_{K}(p)~,
\end{equation}
where we neglected terms that do not depend on $K$. In this case we observe a different behavior of the optimal driving of the energy to the one described in the paragraph above: since the system does not have to reach its limiting probability at $t=\tau$ it can lag behind while the energy reaches its final value. Therefore the optimal protocol will not be overshooting the final energy, but rather quench to it from an intermediary value.

\section{Speed Limit}
In the limit of $\tau$ going to $0$ it is clear from \eref{eq:t_of_p} that it is not possible to realize the transformation of a fixed $p(\tau)$. Therefore there must exist a minimum time $\tau_{min}>0$ for the system to reach $p(\tau)$ from a given $p(0)\neq p(\tau)$, it is determined by $\tau_{min} = \min_{K} F_{K}(p(\tau))$.  It is straightforward to verify that \eref{eq:t_of_p} is monotonically decreasing as a function of $K$. Therefore we can obtain the minimum amount of time required to realize the transformation by taking the limit $K\rightarrow\infty$ and solving the integral
\begin{equation}
	\tau_{min} = 2\int_{p(0)}^{p(\tau)}\!\!dp\frac{ 1}{1-(1+r)p \pm \left[1-(1-r)p\right]}~.
\end{equation}
Which leads to
\begin{equation}\label{eq:speed_limit}
	\tau_{min} = 
	\begin{cases}
		\frac{1}{n\gamma}\log\frac{1-p(0)}{1-p(\tau)}~, & \dot p > 0~,\\
		\frac{1}{m\gamma}\log\frac{p(0)}{p(\tau)}~, &\dot p < 0~,\\
	\end{cases}
\end{equation}
where we explicitly re-introduced the time units $n\gamma$. It is worth noting that these speed limits depend only on the degeneracy of the sector being filled and not on the degeneracy of the sector being emptied: when we are cooling the system ($\dot p < 0$) the dependence on $n$ disappears and, conversely, the dependence on $m$ drops when we are heating the system ($\dot p > 0$)\footnote{ This speed limit is valid within the Markovian regime, a universally valid version of this bound can be derived with the results of \cite{Abiuso2025}.}. In the limit where we achieve the fastest possible protocol ($K\rightarrow\infty$) we can also compute the optimal probability trajectory\footnote{Note that the fastest protocol is also optimal in energy consumption for fixed $\tau$, as it is the only protocol that realizes the transformation in that given amount of time $\tau$.}. We obtain that the fastest we can cool/heat the system is
\begin{equation}
	p(t) = 
	\begin{cases}
		p(0)e^{-n\gamma t}+1-e^{-n\gamma t}~, & \dot p > 0~,\\
		p(0)e^{-m\gamma t}~, &\dot p < 0~.\\
	\end{cases}
\end{equation}
However, it is worth noting that this limit is only realized for a diverging control of the energy $E_\infty(p) = \mp \infty$ and therefore also has a diverging thermodynamic cost.

By letting go of the boundary condition on the state we can use the fact that at short times the state barely evolves $p(t) = p(0) + \dot p(0)t + \mathcal O (t^2)$. Since the curve of the optimal energy control is as function of the probability we obtain that for fast protocol, the optimal curve is almost a constant (regardless of the boundary conditions). Therefore, at first order, the optimal protocol will consist of a large quench at the start, then keeping the energy at an intermediary value from $t=0^+$ to $t=\tau^-$, and finally a quench to its final value. We thus re-obtain, in this specific context, the ``bang-bang'' protocols of \cite{Blaber2021,Rolandi2023Fast}.

\section{Very Degenerate Systems}
In the limit where the degeneracy of one sector is much larger than the other (e.g. \cite{Correa2015,Mehboudi2022,Rolandi2023Collective,Abiuso2024}), we can obtain some simplifications of the results presented above, which lead to interesting dynamics. The evolution of the system towards equilibrium is given by  \eref{eq:to_equilibrium}
\begin{equation}
	\dot p(t) =
	\begin{cases}
		n\gamma f(E(t))(p_{eq}-p(t))~, & r\ll1~,\\
		m\gamma\left[1-f(E(t))\right](p_{eq}-p(t))~, & r\gg1~.
	\end{cases} 
\end{equation}
We can notice that these two regimes behave in the same way under the transformation $n\leftrightarrow m$, $E\rightarrow -E$, and $p\rightarrow 1-p$. Therefore we will focus on the case $r\ll 1$. 

\begin{figure*}[ht]
	\centering
	\includegraphics[width=\textwidth]{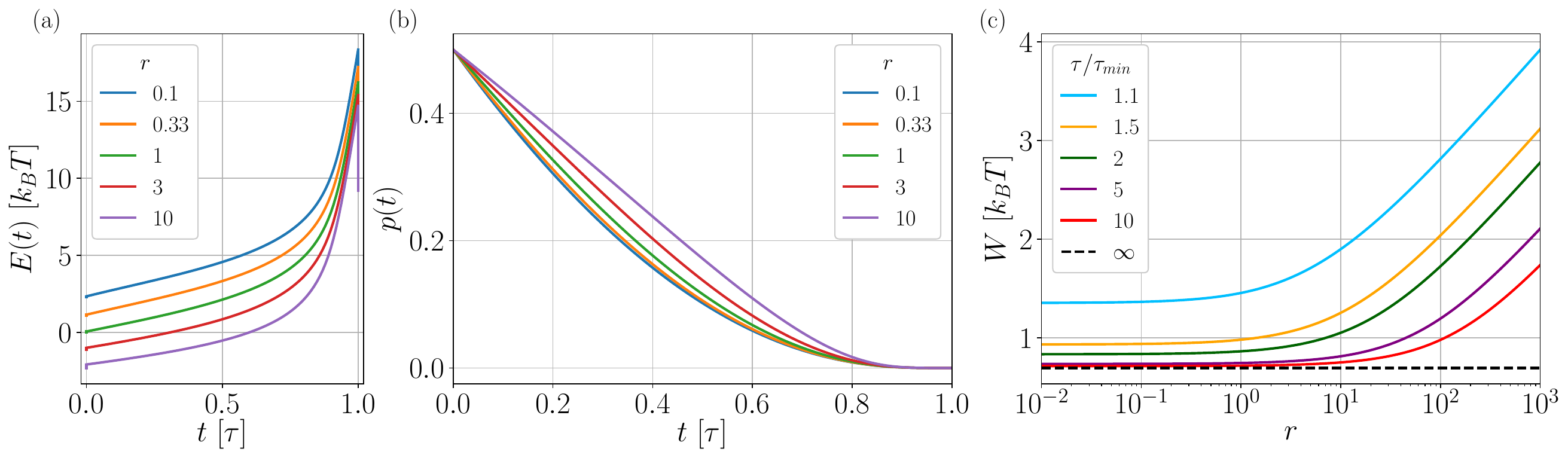}
	\vspace*{-20pt}
	\caption{Optimal finite-time bit erasure protocols ($p(\tau) = 10^{-5}$) for different values of $r$. (a,b) Optimal energy and probability trajectory for finite time erasure protocols ($\tau = 5\tau_{min}$) for different values of $r$. (c) Minimal finite-time erasure cost as a function of $r$ for different protocol durations.}
	\vspace*{-10pt}
	\label{fig:landauer}
\end{figure*}

It is worth noting that in this limit the timescale of relaxation is given by $n\gamma f(E)$. Since $n\gg1$, it is generally large. However it can be exponentially suppressed by taking a gap of the order $-k_BT\ln(r)$, leading to a critical slowdown of the dynamics in this regime.

By expanding \eref{eq:t_of_p} to the leading order in $r$ we find
\begin{equation}\label{eq:degen_int}
	F_K(p(\tau)) = \tau_{min} \pm \frac{1}{K}(p(\tau)-p(0))+\mathcal{O}(r)~.
\end{equation}
Note that $\tau_{min}$ contains a $\mathcal{O}(r^{-1})$ term for $\dot p<0$. Therefore, for fixed boundary conditions on the probability we can solve \eref{eq:K_opt}
\begin{equation}\label{eq:K_degen}
	\kappa_\tau = \frac{|p(\tau)-p(0)|}{\tau -\tau_{min}}~.
\end{equation}
Let us consider the scaling of $\kappa_\tau$ with respect to $r$. Despite the fact that, in principle, $\tau$ is a free parameter, as we will argue in the next section, we should pick it to be a multiple of $\tau_{min}$. Therefore, we find that for heating protocols ($\dot p>0$) $\kappa_\tau = \mathcal{O}(1)$, while for cooling protocols ($\dot p<0$) we have $\kappa_\tau = \mathcal{O}(r)$. Now, we can invert \eref{eq:degen_int} to obtain the optimal protocol as an explicit function of time
\begin{equation}
	p(t) = \begin{cases}
		1-\kappa_\tau w\!\left(\frac{1-p(0)}{\kappa_\tau}e^{\frac{1-p(0)}{\kappa_\tau}-t}\right), & \dot p > 0~,\\
		\frac{\kappa_\tau}{r} w\!\left(\frac{rp(0)}{\kappa_\tau}e^{\frac{rp(0)}{\kappa_\tau}-rt}\right), &\dot p < 0~,\\
	\end{cases}
\end{equation}
for $w(z)$ denoting the principal solution of $z = xe^x$.

Finally, we note that at leading order the optimal energy curve is given by
\begin{equation}
	E_{\kappa_\tau}(p) = \begin{cases}
		k_B T \ln\!\left[\frac{1-p}{\kappa_\tau}\right] + \mathcal{O}(r), & \dot p> 0~,\vspace*{2pt}\\
		k_B T \ln\!\left[\frac{\kappa_\tau(1-p)}{r^2p^2}\right] + \mathcal{O}(r), & \dot p< 0~.
		\end{cases}
\end{equation}
This equation can be easily integrated analytically to obtain the corresponding minimal amount of work with \eref{eq:work_opt}. We observe that for cooling protocols the work cost diverges as $\sim\log(r)$.

\section{Bit Erasure}
We will now showcase the difference in the optimal protocols for different values of $r$ with the paradigmatic task of erasing a bit of information in finite time. This allows us to benchmark the quantitative and qualitative differences between the effective two-level systems with different degeneracies. Bit erasure is defined by the boundary conditions $p(0) = \frac{1}{2}$ and $p(\tau) \approx 0$\footnote{We require only approximate erasure, as perfect erasure in finite time is impossible~\cite{Goold2015}. In this case, we can see that \eref{eq:speed_limit} diverges for $p(\tau) = 0$.}~\cite{Landauer1961}. For the system to be at equilibrium at the start and the end of the protocol we have $E(0) = -k_B T\ln r$ and $E(\tau) = k_BT\ln\frac{1-p(\tau)}{rp(\tau)}$. In order to have a fair comparison between systems with different degeneracy ratios, we would want all the relaxation timescales to be identical. However, \eref{eq:to_equilibrium} shows that the relaxation timescale depends on the energy gap (except for $r=1$). Therefore, we cannot ensure that the relaxation timescale remains the same between systems with different degeneracies while the gap is being driven. It is worth noting that in general relaxation time-scales depend on the energetic structure of the system, and therefore one has to take extra care when considering a driven system. In this case the quantity that can be compared between systems is the speed limit to realize the task, given by \eref{eq:speed_limit}: $\tau_{min} = r^{-1}\log\frac{p(0)}{p(\tau)}$. Indeed, we cannot compare the finite-time erasure of systems with different degeneracy ratios if for some values of $r$ the task cannot even be realized in that duration of time. Therefore, here we compare the finite-time erasure protocols and their energetic cost while varying $r$ and keeping $\tau/\tau_{min}$ fixed.

We showcase the obtained optimal protocols $E(t)$ for finite-time erasure ($p(\tau) = 10^{-5}$) for a relatively slowly driven protocol ($\tau = 5\tau_{min}$ which is slow but not slow enough to fully be in the quasistatic regime) in \sfref{fig:landauer}{a} and their corresponding state trajectories $p(t)$ in \sfref{fig:landauer}{b}. Similarly to the findings of \cite{Esposito2010,Aggarwal2025} we observe quenches at the start and end of the protocol for the optimal energy control.
Let us first note that the shift in the extremal points of the energy protocols (and thus in the bulk) is due to the fact that we are imposing boundary conditions on the probabilities. From \eref{eq:equilibrium}, we can see that to keep the same equilibrium probability, a change in $r$ needs to be compensated by a shift in the energy gap. However, even after correcting for the energy shift, we can observe differences between the protocols, and these lead to different probability trajectories. Ultimately, these differences result in a dependence of the minimal energetic cost of erasure on the degeneracy ratio, as we show in \sfref{fig:landauer}{c}. Each line corresponds to a different ratio of protocol time to the speed limit. As expected, the cost is lower for slower protocols and is ultimately bounded by the Landauer bound
$W\geq k_BT\ln2$~\cite{Landauer1961} which we show with the dashed black line as it is achieved for $\tau\rightarrow\infty$.\\

The implementation of the optimization presented in the previous sections and the code used to obtain the results in this section are publicly available~\cite{code}.

\section{Conclusion}

In this work, we investigated how internal degeneracies in the energy levels of a two-level system influence minimally dissipating thermodynamic operations. Extending the framework of Ref.~\cite{Esposito2010}, we considered two-level systems with internal degeneracies and analyzed the optimization of these \emph{effective} two-level systems, which play a central role in multiple frameworks ranging from theoretical metrology~\cite{Scandi2025,Correa2015,Mehboudi2022,Abiuso2024} to experimental quantum technologies~\cite{Myers2022,roadmap2025}. We derived closed-form analytical expressions for the minimally dissipating state trajectories and control protocols in terms of the degeneracy ratio between the two levels and provided explicit formulas for the corresponding work cost. The performance of these optimal protocols was further illustrated and benchmarked in the context of bit erasure. Starting from a microscopic description, we demonstrated that this optimization applies to any system that can be coarse-grained into a (Markovian) effective two-level description, while still allowing for internal quantum dynamics within each effective level. We further showed that degeneracies alter the dynamical and equilibrium properties of the system in non-trivial ways. In particular, the degeneracy of the level being populated during a process imposes a fundamental bound on the achievable system evolution.

Looking ahead, these results provide a blueprint to implement minimally dissipating protocols in both classical and quantum two-level systems, enabling a wide range of applications. For example, theoretically, they can be used to design resource-optimal, ultra-precise thermometers. Experimentally, they enable thermodynamically optimal operations in mesoscopic systems such as quantum dots. Moreover, this analysis can be exploited to characterize the presence of degeneracies in the system and benchmark the degree of control over experimental platforms. \\

\emph{Acknowledgments.} The author warmly thanks Martí Perarnau-Llobet, Mark T. Mitchison, Maximilian Lock, and Pharnam Bakhshinezhad for insightful discussions. The author acknowledges support from the the Swiss National Science Foundation for funding through Postdoc.Mobility (Grant No. P500PT225461).

\bibliographystyle{apsrev4-2}
\bibliography{my_bib.bib}


\end{document}